\documentclass{svproc}
\usepackage{url,bm}

\begin{document}

\mainmatter              
\title{Theory of Inclusive B Decays}
\titlerunning{Theory of Inclusive B Decays}  
%
\author{Gil Paz}
\authorrunning{Gil Paz} 
%
\tocauthor{Gil Paz}
\institute{Department of Physics and Astronomy,\\
Wayne State University, \\
Detroit, Michigan 48201, USA}

\maketitle              

\begin{abstract}
In this talk I briefly review the current status of the theory of the inclusive $B$ decays: $\bar B\to X_c\, \ell\,\bar\nu_\ell$, $\bar B\to X_u\, \ell\,\bar\nu_\ell$, and $\bar B\to X_s\,\gamma$. I try to answer three questions: what is the current ``state of the art", can the theoretical prediction be improved, and will it lead to a smaller theoretical uncertainty.
\end{abstract}
\section{Introduction}
Why study inclusive $B$ decays? First, they are part of flavor physics which allows access to new physics at scales beyond the reach of current colliders. For example, $K-\bar K$ mixing and  $B-\bar B$ mixing probe scales above hundreds of TeV \cite{UTfit:2007eik,Bevan:2014cya}.  Second, there is a persistent tension between the extraction of  $|V_{cb}|$ and $|V_{ub}|$ from inclusive $B$ decays on the one hand and the extraction of $|V_{cb}|$ and $|V_{ub}|$ from exclusive $B$ decays on the other. The inclusive values are consistently larger than the exclusive ones, see every biennial edition of the Review of Particle Physics from 2006 \cite{ParticleDataGroup:2006fqo} to 2022 \cite{ParticleDataGroup:2022pth}. Third, they test basic QFT tools such as factorization theorems and the Operator Product Expansion (OPE) to higher orders both perturbatively and non-perturbatively. For example, the OPE  for $\bar B\to X_c\, \ell\,\bar\nu_\ell$ is now known perturbatively  to third order and non-perturbative to fourth order, see for example the talks of Keri Vos and Matteo Fael at \cite{Barolo}. Fourth, they are a ``window" to non-perturbative physics. For example, at leading twist the $\bar B\to X_s\,\gamma$ photon spectrum \emph{is} the $B$-meson $b$-quark pdf. 

How do we make theoretical predictions? At energies below $m_W, m_Z$ and $m_t$ the effective Hamiltonian is known, see \cite{Buras:1998raa}. For example, for $\bar B\to X_s\,\gamma$ it is 
\begin{equation}
{\cal H}_{\mbox{\scriptsize eff}}=\frac{G_F}{\sqrt{2}}\sum_{q=u,c}V_{qb}V_{qs}^{*}\left( C_1 Q_1^q+C_2 Q_2^q+\!\!\sum_{i=3}^{10}C_iQ_i+C_{7\gamma}Q_{7\gamma}+C_{8g}Q_{8g}\right)+\mbox{h.c.}\,.
\end{equation}
The coefficients $C_i$ are calculable in perturbation theory while $Q_i$ are operators with non-perturbative matrix elements.  In particular, 
\begin{eqnarray}\label{operators}
Q_1^q&=&(\bar q b)_{V-A}(\bar s q)_{V-A}\quad (q=u,c)\nonumber\\
Q_{7\gamma}&=&\frac{-e}{8\pi^2}m_b\bar s
\sigma_{\mu\nu}(1+\gamma_5)F^{\mu\nu}b\nonumber\\
Q_{8g}&=&\frac{-g_s}{8\pi^2}m_b\bar s
\sigma_{\mu\nu}(1+\gamma_5)G^{\mu\nu}b.
\end{eqnarray}
The main problem is that we know the operators but usually we cannot calculate the matrix elements. This is because the operators contain strongly interacting quarks and gluons. These operators can be \emph{local}, e.g.,  $\bar{q}(0)\cdots q(0)$, or non-local, e.g., $\bar{q}(0)\cdots q(tn)$, where $n$ is a light-like vector. In particular, for inclusive $B$ decays we encounter ``diagonal" matrix elements between $B$-meson states. The operator can be local, as in   $\langle {\bar B}|\bar b\, \bm{\vec{D}^2}\,b|{\bar B}\rangle=2M_B{\mu_{\pi}^2}$, or non local as in $\int_{-\infty}^{\infty}\,dt\,e^{i\omega t}\langle \bar{B}(v)|\bar{h}(0)\,[0,tn]\,h(tn)|\bar{B}(v)\rangle=(2\pi)(2M_B)S(\omega)$, where $h$ is a heavy quark and $[0,tn]$ a Wilson line.

What to do with the non-perturbative objects? There are several possible strategies. First, if possible, we can extract them \emph{carefully} from data. For example, the parameter ${\mu_{\pi}^2}$ is extracted  from $\bar B\to X_c\, \ell\,\bar\nu_\ell$. Second, sometimes we can calculate them using some non-perturbative method, e.g. Lattice QCD. For example, the $B$-meson decay constant is calculated this way. Third, we can use symmetries to reduce the number of non-perturbative objects. For example, use $SU(3)$ flavor to calculate exclusive hadronic decays of $B$ mesons. Fourth, when all else fails, use conservative modeling. For example, this is how the non-perturbative error for $\bar B\to X_s\,\gamma$ is determined. 

An important feature of inclusive $B$ decays is that we have two expansion parameters we can use. A perturbative parameter: $\alpha_s$ which at the scale $m_b$ is about $0.2$ and a non-perturbative parameter: $\Lambda_{\mbox{\scriptsize QCD}}/{m_b}$ which is about $0.1$. Thus observables can be calculated as a double series in these parameters. 

How well can we calculate? Answering this question is the goal of this talk. In the following I will try to answer three questions. What is the current ``state of the art"? Can the theoretical prediction be improved? Will it lead to a smaller theoretical uncertainty? I will consider three topics: $|V_{cb}|$ and $\bar B\to X_c\, \ell\,\bar\nu_\ell$ in section \ref{b2c},  $|V_{ub}|$ and $\bar B\to X_u\, \ell\,\bar\nu_\ell$ in section \ref{b2u}, and $\bar B\to X_s\,\gamma$ in section \ref{b2s}. For more details on the current status, see also the talks at the ``Challenges in Semileptonic B Decays" Workshop held in April 2022 in Barolo, Italy \cite{Barolo}.

\section{$|V_{cb}|$ and $\bar B\to X_c\, \ell\,\bar\nu_\ell$}\label{b2c}
Inclusive semi-leptonic  $b\to c$ decays, namely, $\bar B\to X_c\, \ell\,\bar\nu_\ell$, are governed by the Hamiltonian,
\begin{equation}
{\cal H}_{\mbox{\scriptsize eff}}=\frac{G_F}{\sqrt{2}}C_1(\mu)V_{cb}\,\bar \ell\gamma_\mu(1-\gamma^5)\nu_\ell\,\bar c \gamma^\mu(1-\gamma^5)b. 
\end{equation}
Using the optical theorem we can calculate $\bar B\to X_c\, \ell\,\bar\nu_\ell$ as an OPE. Up to order $1/m_b^2$ we have symbolically $\Gamma \sim c_0 \langle O_0\rangle+c_2^{\,j} \langle O_2^j\rangle{/m_b^2}+\cdots$, with summation over $j$ implied. In this expression $c_0 \langle O_0\rangle$ is a free quark decay. At tree level it is the  same as $\mu\to e\, \bar\nu_e \nu_\mu$.  The coefficients $c_i^{\, j}$ are perturbative and can be calculated as a series in $\alpha_s$. The matrix elements $\langle O_i^j \rangle $ are non perturbative and can be  extracted from experiment. For example, $\langle O_0\rangle=\langle \bar B|{ \bar b b}| \bar B\rangle=2M_B$, $\langle O_2^{\mbox{\scriptsize kin.}}\rangle=\langle \bar B|{ \bar b (iD)^2 b}| \bar B\rangle$ is related to the HQET parameter  $\mu_\pi^2$, and $\langle O_2^{\mbox{\scriptsize mag.}}\rangle=\langle \bar B|{ \bar b\, \sigma_{\mu\nu}G^{\mu\nu}b}| \bar B\rangle$ is related to the HQET parameter $\mu_G^2$ and can be extracted from the $B^*-B$ mass difference. 

The current ``state of the art" is that corrections up to and including $1/m_b^5$ are known. Considering the Wilson coefficients at tree level we have: at $1/m_b^0$ one operator, at $1/m_b$ zero operators, at $1/m_b^2$ two operators \cite{Blok:1993va,Manohar:1993qn},  at $1/m_b^3$ two operators \cite{Gremm:1996df}, at $1/m_b^4$ nine operators, and at $1/m_b^5$ eighteen operators \cite{Mannel:2010wj}. Are these all the possible operators at each order in $1/m_b$?

This question was answered in \cite{Gunawardana:2017zix}, where a method was presented that can list such bilinear operators, in principle, to \emph{arbitrary} dimension. It also gives NRQED and NRQCD bilinear operators to arbitrary dimension. See also the work of \cite{Kobach:2017xkw} using Hilbert series. It turns out that the counting above does not include all the possible operators. Allowing for Wilson coefficients at order $\alpha_s$ or higher, at $1/m_b^4$ there are eleven operators instead of nine,  and at $1/m_b^5$ there are twenty five operators instead of eighteen.  The contribution of these extra operators to $\bar B\to X_c\, \ell\,\bar\nu_\ell$ is unknown but extremely small. For example, a $1/m_b^4$ operator with an $\alpha_s$ coefficient would give a relative contribution of  $\alpha_s\left(\Lambda_{\mbox{\scriptsize QCD}}/m_b\right)^4\sim 0.2\cdot (0.1)^4\sim 10^{-5}$. The $1/m_b^4, 1/m_b^5$ matrix elements with tree level Wilson coefficients were extracted from $\bar B\to X_c\ell\bar\nu_\ell$ in \cite{Gambino:2016jkc}. The authors of \cite{Gambino:2016jkc} find that ``The higher power corrections have a minor effect on $|V_{cb}|$ ... There is a $-0.25\%$ reduction in $|V_{cb}|$''.

What is the current ``state of the art"?  As of 2021
\begin{equation}\label{OPE}
\Gamma \sim c_0 \langle O_0\rangle+c_2^{\,j} \frac{\langle O_2^j\rangle}{m_b^2}+c_3^{\,j} \frac{\langle O_3^j\rangle}{m_b^3}+c_4^{\,j} \frac{\langle O_4^j\rangle}{m_b^4}+c_5^{\,j} \frac{\langle O_5^j\rangle}{m_b^5}+\cdots,
\end{equation}
where $c_0$ is known at ${\cal O}(\alpha_s^0)$, at  ${\cal O}(\alpha_s^1)$ \cite{Trott:2004xc,Aquila:2005hq}, at ${\cal O}(\alpha_s^2)$ \cite{Melnikov:2008qs,Pak:2008cp}, and at ${\cal O}(\alpha_s^3)$ \cite{Fael:2020tow} for the total rate, $c^j_2$ are known at ${\cal O}(\alpha_s^0)$ \cite{Blok:1993va,Manohar:1993qn} and at ${\cal O}(\alpha_s^1)$ \cite{Becher:2007tk,Alberti:2012dn,Alberti:2013kxa,Mannel:2015jka}$, c^j_3$ are known at ${\cal O}(\alpha_s^0)$ \cite{Gremm:1996df}, and at ${\cal O}(\alpha_s^1)$ \cite{Mannel:2019qel,Mannel:2021zzr} for selected observables,  
and $c^j_{4}$ and $c^j_{5}$ are known at ${\cal O}(\alpha_s^0)$ \cite{Mannel:2010wj}. The state of the art value of inclusive $|V_{cb}|$ is $42.16(51)\cdot 10^{-3}$ \cite{Bordone:2021oof}, while the 2022 Review of Particle Physics lists the value of exclusive $|V_{cb}|$ as $39.4(8)\cdot 10^{-3}$ \cite{ParticleDataGroup:2022pth}.  The exclusive/inclusive $|V_{cb}|$ puzzle remains.  

Can the theoretical prediction be improved? Yes, for example $c^j_3$ at ${\cal O}(\alpha_s^1)$ will probably be calculated in the coming years for the fully differential decay rate. Will it lead to a smaller theoretical uncertainty?  Probably. There could be other improvements. For example, the use of the leptonic invariant mass spectrum that depends on less parameters \cite{Fael:2018vsp}. This method was very recently implemented in \cite{Bernlochner:2022ucr} finding $|V_{cb}|=41.69(63)\cdot 10^{-3}$.

\section{$|V_{ub}|$ and $\bar B\to X_u\, \ell\,\bar\nu_\ell$}\label{b2u}
$|V_{ub}|$ plays an important role in the unitarity triangle fit. Like $|V_{cb}|$, $|V_{ub}|$ inclusive  is larger than $|V_{ub}|$ exclusive. The 2022 Review of Particle Physics lists for inclusive $|V_{ub}|$ the value $(4.13\pm 0.12^{+0.13}_{-0.14}\pm0.18)\cdot 10^{-3}$ and for exclusive $|V_{ub}|$ the value $(3.70\pm 0.10\pm0.12)\cdot 10^{-3}$ \cite{ParticleDataGroup:2022pth}.

\emph{If} we could measure the total $\Gamma(\bar B\to X_u\, l\,\bar\nu)$ we could use a \emph{local} OPE, just like for $\bar B\to X_c\, \ell\,\bar\nu_\ell$, see equation (\ref{OPE}). Since $\Gamma(\bar B\to X_c\ell\,\bar\nu_\ell)\gg(\bar B\to X_u\ell\,\bar\nu_\ell)$ the total rate \emph{cannot}~be measured and one needs to cut the charm background. For example, by requiring that $M^2_X<M_D^2\sim m_b\,\Lambda_{\mbox{\scriptsize QCD}}$. This is not inclusive enough for a local OPE, but a non-local OPE is still possible. More generally, one can distinguish three regions depending on the size of $M^2_X$. The region $M^2_X\sim m_b^2$ is the "OPE region", where we have a local OPE. The region $M^2_X\sim  m_b\,\Lambda_{\mbox{\scriptsize QCD}}$ is the ``end-point region" where we have a non-local OPE. The region $M^2_X\sim \Lambda^2_{\mbox{\scriptsize QCD}}$ is the ``resonance region'' where no inclusive description is possible and one must use an exclusive description. 

For $\bar B\to X_u\, l\,\bar\nu$ we therefore have a non-local OPE, where the spectra is described in terms on non-local matrix elements called ``shape functions" or ``soft functions". Schematically we have 
\begin{equation}\label{nonlocal}
d\Gamma\sim H\cdot J \otimes S+\frac1{m_b}\sum_i {H}\cdot {J} \otimes{ s_i}+\cdots\,. 
\end{equation}
At leading power in $\Lambda_{\mbox{\scriptsize QCD}}/m_b$, $d\Gamma$ factorizes to a product of a ``hard" ($H$) and ``jet" (${J}$) functions convoluted with a a non-perturbative ``shape function" ($S$) \cite{Neubert:1993ch,Neubert:1993um,Bigi:1993ex}. Intuitively $S$ is the $B$-meson $b$-quark pdf. A similar factorization formula holds for  $\bar B\to X_s\,\gamma$ at the photon energy end-point region. As a result, at leading power in $\Lambda_{\mbox{\scriptsize QCD}}/m_b$, $S$ is the $\bar B\to X_s\,\gamma$ photon spectrum. As an example, see its recent extraction by the SIMBA Collaboration \cite{Bernlochner:2020jlt}.

What about $\Lambda_{\mbox{\scriptsize QCD}}/m_b$ corrections? At this order several subleading shape functions (SSF) appear denoted in equation (\ref{nonlocal}) by $s_i$ \cite{Lee:2004ja,Bosch:2004cb,Beneke:2004in,Bauer:2001mh,Leibovich:2002ys}. They are analogous to ``higher twist" contributions for deep inelastic scattering.  Different linear combinations of SSF appear for $\bar B\to X_u\, \ell\,\bar\nu_\ell$ and $\bar B\to X_s\,\gamma$. Furthermore, $\bar B\to X_s\,\gamma$ has unique SSF, called ``resolved photon contributions", see section \ref{b2s}.

Moments of the shape functions are related to ``universal" matrix elements that appear also for $\bar B\to X_c\, \ell\,\bar\nu_\ell$ . For example, the first moment of $S$ is related to the $b$-quark mass and the second moment is related to $\mu_\pi^2$.

There are different theoretical frameworks for $|V_{ub}|$  extractions. They  use similar perturbative inputs, currently at ${\cal O} (\alpha_s)$, but they differ in how they extract (or model) $S$  and how they treat power corrections. Current  theoretical frameworks used by experiments are designated by acronyms. Thus we have BLNP \cite{Lange:2005yw}, DGE \cite{Andersen:2005mj}, GGOU \cite{Gambino:2007rp}, and ADFR \cite{Aglietti:2007ik}. A recent extraction of inclusive $|V_{ub}|$ from Belle data finds almost no difference between the  $|V_{ub}|$  values from these frameworks \cite{Belle:2021eni}. See also Francesco TenchiniÕs talk at this workshop. We should keep in mind though that these frameworks were developed before 2010.

Can the theoretical prediction be improved? Yes, many NNLO calculations are known. The free quark differential decay rate was calculated at ${\cal O} (\alpha_s^2)$ in \cite{Brucherseifer:2013cu}. The hard function $H$ is known at ${\cal O} (\alpha_s^2)$ \cite{Bonciani:2008wf,Asatrian:2008uk,Beneke:2008ei,Bell:2008ws}. The jet function $J$ was calculated at ${\cal O} (\alpha_s^2)$ in \cite{Becher:2006qw} and at ${\cal O} (\alpha_s^3)$ in \cite{Bruser:2018rad}. Corrections that scale like ${\cal O} (\alpha_s)\times{
\cal O}(\Lambda_{\mbox{\scriptsize QCD}}/m_b)$ in the form of subleading jet function ($j_i$) were calculated in \cite{Paz:2009ut}. Moments of the leading ($S$) and subleading shape functions ($s_i$) can be calculated using the data from \cite{Gambino:2016jkc} and the method of \cite{Gunawardana:2017zix} (see also the appendix of \cite{Heinonen:2016cwm}).  For frameworks that use $\bar B\to X_s\,\gamma$, there is also the calculation of the resolved photon contributions in \cite{Benzke:2010js}. All of these are not fully combined yet.  Will it lead to a smaller theoretical uncertainty?  Not necessarily. 

\section{$\bar B\to X_s\,\gamma$}\label{b2s}
$\bar B\to X_s\,\gamma$ is a well known probe of physics beyond the standard model. The 2022 edition of the Review of Particle Physics lists the CP averaged branching ratio as $(3.49\pm0.19)\cdot10^{-4}$ \cite{ParticleDataGroup:2022pth}. The standard model prediction in 2015 was $(3.36\pm0.23)\cdot 10^{-4}$ \cite{Misiak:2015xwa}. The largest uncertainty of this prediction, about $5\%$, is non-perturbative from ``resolved photons contributions".  These are contributions in which the photon does not couple directly to the weak vertex  \cite{Benzke:2010js}. They are parameterized by fields that are  localized on two different light cones. In some sense they are a precursor to the contributions discussed in Matthias Neubert's talk at this workshop.

At ${\cal O}(\Lambda_{\mbox{\scriptsize QCD}}/m_b)$ the resolved photon contributions arise from the interference of $Q_1-Q_{7\gamma}$, $Q_{7\gamma}-Q_{8g}$ and $Q_{8g}-Q_{8g}$. See equation (\ref{operators}) for their definitions. 
The standard model CP asymmetry is dominated by $Q_1^q-Q_{7\gamma}: -0.6\%<{\cal A}_{X_s\gamma}^{\mbox{\scriptsize SM}}<2.8\%$ \cite{Benzke:2010tq}, while the 2022 edition of the Review of Particle Physics lists the range from experiment as ${\cal A}_{X_s\gamma}=1.5\% \pm1.1\%$ \cite{ParticleDataGroup:2022pth}.

Can these be improved? The $Q_{8g}-Q_{8g}$ contribution is hard to improve, but beyond the power suppression it is suppressed  by the square of the $s$-quark charge. The $Q_{7\gamma}-Q_{8g}$ contribution is constrained by isospin asymmetry $\bar B^{0/\pm}\to X_s\,\gamma$ \cite{Misiak:2009nr}. This uncertainty was reduced by a Belle measurement \cite{Belle:2018iff}.  The $Q_1-Q_{7\gamma}$ contribution depends on a non-perturbative function $g_{17}(\omega,\omega_1)$  whose moments can be extracted from $\bar B\to X_c\, \ell\,\bar\nu_\ell$ OPE. The 2010 analysis of \cite{Benzke:2010js} only had two non-zero moments. A 2019 analysis \cite{Gunawardana:2019gep} by Ayesh Gunawardana and me added  six non-zero moments using the data from \cite{Gambino:2016jkc} and the methods of \cite{Gunawardana:2017zix}. Using moments we can better model the $Q_1-Q_{7\gamma}$ resolved photon contribution. The new estimate reduced the uncertainty on the  total rate by up to $50\%$, but increased the uncertainty on the CP asymmetry by about $33\%$.  These improvements were included in the 2020 updated standard model prediction of $(3.40\pm0.17)\cdot 10^{-4}$ \cite{Misiak:2020vlo}. Using different models, including some ${\cal O}\left({\Lambda^2_{\mbox{\scriptsize QCD}}}/{m^2_b}\right)$ corrections and larger $m_c$ range, a smaller reduction was found in \cite{Benzke:2020htm}.

Can the theoretical prediction be improved? Yes, the $m_c$ dependence can only be really controlled by an NLO analysis of the $Q_1-Q_{7\gamma}$ contribution.  Will it lead to a smaller theoretical uncertainty?  Not necessarily. 

\section{Conclusions}
Flavor physics probes very high scales and advanced theoretical tools. This decade will be very exciting with, e.g., LHCb and Belle II data. These supply puzzles and tensions that motivate further theoretical work. A big theoretical challenge is controlling non-perturbative effects. I  discussed the ``state of the art" of $|V_{cb}|$ and $\bar B\to X_c\, \ell\,\bar\nu_\ell$, 
$|V_{ub}|$ and $\bar B\to X_u\, \ell\,\bar\nu_\ell$, and $\bar B\to X_s\,\gamma$. While the theory is fairly advanced, there is still room for theoretical improvements. These will lead to a more robust  theoretical uncertainty but not necessarily to a smaller one.

\section*{Acknowledgements} I thank the organizers of the 8th Workshop on Theory, Phenomenology and Experiments in Flavour Physics: Neutrinos, Flavor Physics and Beyond (FPCapri22) for the invitation to give this talk. This work was supported in part by the DOE grant DE-SC0007983.

%
%

\end{document}